\documentclass[12pt]{article}
\newlength{\dinwidth}
\newlength{\dinmargin}
\usepackage{graphicx}          

\setkeys{Gin}{keepaspectratio} 

\begin{document}
\newcommand{\f}{$F_2^{\gamma}$}
\newcommand {\pom}  {I\hspace{-0.2em}P}

\def\ap#1#2#3   {{\em Ann. Phys. (NY)} {\bf#1} (#2) #3.}   
\def\apj#1#2#3  {{\em Astrophys. J.} {\bf#1} (#2) #3.} 
\def\apjl#1#2#3 {{\em Astrophys. J. Lett.} {\bf#1} (#2) #3.}
\def\app#1#2#3  {{\em Acta. Phys. Pol.} {\bf#1} (#2) #3.}
\def\ar#1#2#3   {{\em Ann. Rev. Nucl. Part. Sci.} {\bf#1} (#2) #3.}
\def\cpc#1#2#3  {{\em Computer Phys. Comm.} {\bf#1} (#2) #3.}
\def\err#1#2#3  {{\it Erratum} {\bf#1} (#2) #3.}
\def\ib#1#2#3   {{\it ibid.} {\bf#1} (#2) #3.}
\def\jmp#1#2#3  {{\em J. Math. Phys.} {\bf#1} (#2) #3.}
\def\ijmp#1#2#3 {{\em Int. J. Mod. Phys.} {\bf#1} (#2) #3}
\def\jetp#1#2#3 {{\em JETP Lett.} {\bf#1} (#2) #3.}
\def\jpg#1#2#3  {{\em J. Phys. G.} {\bf#1} (#2) #3.}
\def\mpl#1#2#3  {{\em Mod. Phys. Lett.} {\bf#1} (#2) #3.}
\def\nat#1#2#3  {{\em Nature (London)} {\bf#1} (#2) #3.}
\def\nc#1#2#3   {{\em Nuovo Cim.} {\bf#1} (#2) #3.}
\def\nim#1#2#3  {{\em Nucl. Instr. Meth.} {\bf#1} (#2) #3.}
\def\np#1#2#3   {{\em Nucl. Phys.} {\bf#1} (#2) #3}
\def\pcps#1#2#3 {{\em Proc. Cam. Phil. Soc.} {\bf#1} (#2) #3.}
\def\pl#1#2#3   {{\em Phys. Lett.} {\bf#1} (#2) #3}
\def\prep#1#2#3 {{\em Phys. Rep.} {\bf#1} (#2) #3}
\def\prev#1#2#3 {{\em Phys. Rev.} {\bf#1} (#2) #3}
\def\prl#1#2#3  {{\em Phys. Rev. Lett.} {\bf#1} (#2) #3}
\def\prs#1#2#3  {{\em Proc. Roy. Soc.} {\bf#1} (#2) #3.}
\def\ptp#1#2#3  {{\em Prog. Th. Phys.} {\bf#1} (#2) #3.}
\def\ps#1#2#3   {{\em Physica Scripta} {\bf#1} (#2) #3.}
\def\rmp#1#2#3  {{\em Rev. Mod. Phys.} {\bf#1} (#2) #3}
\def\rpp#1#2#3  {{\em Rep. Prog. Phys.} {\bf#1} (#2) #3.}
\def\sjnp#1#2#3 {{\em Sov. J. Nucl. Phys.} {\bf#1} (#2) #3}
\def\spj#1#2#3  {{\em Sov. Phys. JEPT} {\bf#1} (#2) #3}
\def\spu#1#2#3  {{\em Sov. Phys.-Usp.} {\bf#1} (#2) #3.}
\def\zp#1#2#3   {{\em Zeit. Phys.} {\bf#1} (#2) #3}

\title{\vspace{5cm}
\bf{ The ALLM parameterization of $\sigma_{tot}(\gamma^* p)$\\ an update
}
\vspace{2cm}}

\author{
 {\bf H.~Abramowicz and A.~Levy} \\ 
{\small \sl School of Physics and Astronomy}\\ {\small \sl Raymond and 
Beverly Sackler Faculty of Exact Sciences}\\
  {\small \sl Tel--Aviv University, Tel--Aviv, Israel}
}
\date{ }
\maketitle

\begin{abstract}

The ALLM parameterization of $\sigma_{tot}(\gamma^* p)$ has been
updated by using all published $F_2$ data to determine its
parameters. The fit yields a $\chi^2$/ndf of 0.97 for the 1356 data
points. The updated ALLM parameterization, ALLM97, gives a good
description of all the available data in the whole $x$ and $Q^2$ range
studied so far ($3\times 10^{-6} < x < 0.85,\ \ 0
\le Q^2 < 5000$ GeV$^2$).

\end{abstract}

\vspace{-17cm}
\begin{flushleft}
DESY 97--251 \\
December 1997 \\
\end{flushleft}

\setcounter{page}{0}
\thispagestyle{empty}
\newpage  

\section{Introduction}

The ALLM (Abramowicz, Levin, Levy, Maor) parameterization~\cite{allm}
is a parameterization for describing the total $\gamma^* p$ cross
section, $\sigma_{tot}(\gamma^* p)$, above the resonance region in the
whole $Q^2$ range, where $Q^2$ is the negative of the four-momentum
transfer squared of the exchanged photon in $ep$ interactions. It is
constructed so as to include also the real photon ($Q^2=0$) cross
section. There were two main reasons for such a parameterization. The
practical reason was that it was needed for estimating acceptance
corrections and radiative corrections in the whole $Q^2$ region for
$W^2>$ 3 GeV$^2$, where $W$ is the $\gamma^* p$ center of mass
energy. The theoretical reason was the hope that it would shed light on
the transition region between the soft and hard interactions and their
interplay.

\begin{figure}[tbh]
\begin{center}
  \includegraphics [bb=41 164 511 645,width=\hsize,totalheight=20cm]
  {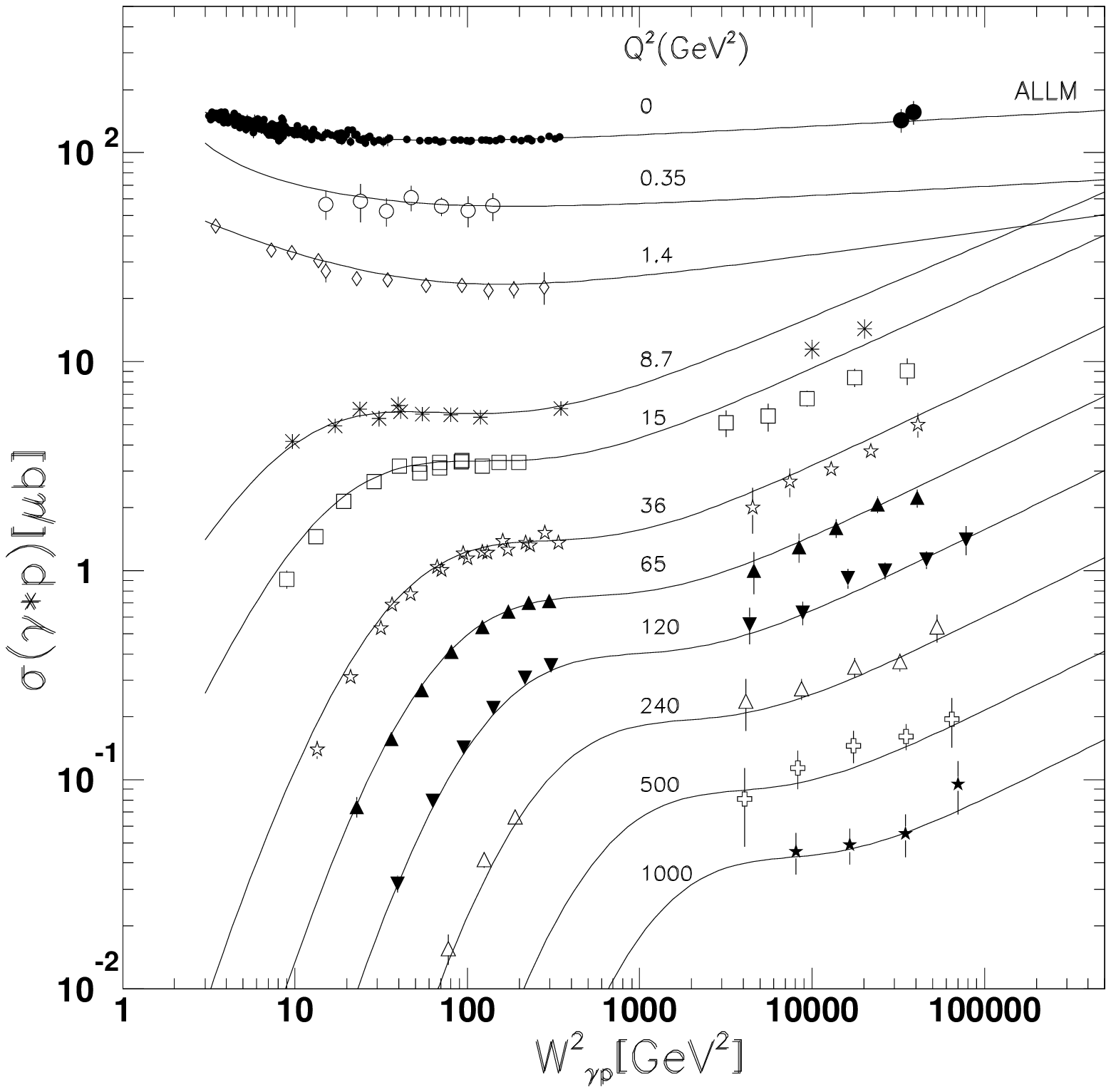}
\end{center}
\vspace{-.5cm}
\caption
 {
  The total $\gamma^* p$ cross section as function of $W^2$, 
for different $Q^2$ values. The curves are the expectations of the ALLM
parameterization.
  }
\label{fig:allm-95}
\end{figure}

The parameterization is based on a Regge motivated approach, similar
to that used earlier by Donnachie and Landshoff~\cite{dlq2}, extended
into the large $Q^2$ regime in a way compatible with QCD
expectations. The data used to fit the parameters were all the $F_2$
measurements available in 1991 together with the total photoproduction
cross section data, which were measured at that time up to center of
mass energies of $W \approx$ 20 GeV. In spite of the fact that the fit
used relatively high Bjorken $x$ data and data of $\sigma_{tot}(\gamma
p)$ at low energies, its predictions agreed amazingly well with the
new HERA data, both for real and virtual photon cross sections. This
can be seen in figure~\ref{fig:allm-95} which shows the total
$\gamma^* p$ cross section as function of the center of mass energy
squared, $W^2$, for fixed $Q^2$ values~\cite{al95}. The curves are the
ALLM parameterization which were fitted to the lower energy data ($W^2
< 400$ GeV$^2$) and extrapolated to the first HERA measurements. The
predictions did very well for the photoproduction data as well as for
the $Q^2>30$ GeV$^2$ region. However, at low $x$, in the intermediate
5 $< Q^2 <$ 25 GeV$^2$ region the predictions turned out to be higher
than the data.

\begin{figure}[tbh]
\begin{center}
  \includegraphics [bb=18 68 547 744,width=\hsize,totalheight=20cm]
  {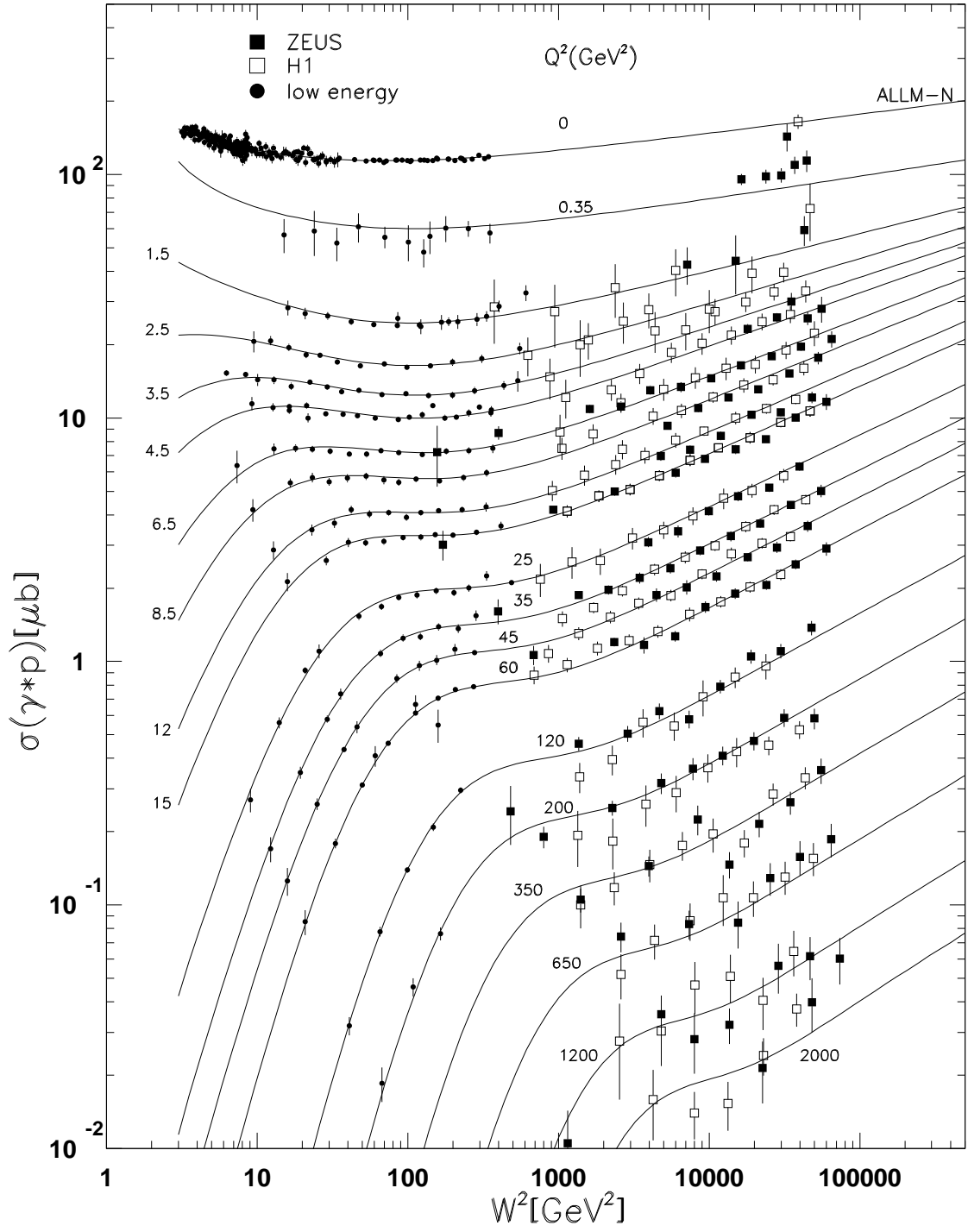}
\end{center}
\vspace{-.5cm}
\caption
 {
  The total $\gamma^* p$ cross section as function of $W^2$, 
for different $Q^2$ values. The curves are the results of the ALLM-N
parameterization.
  }
\label{fig:allm-n}
\end{figure}

An attempt was made in 1995 to include the first HERA $F_2$ data in
the fit. The new parameterization, ALLM-N~\cite{alon}, did somewhat
better in the region of $Q^2 \approx$ 10 GeV$^2$, as can be seen in
figure~\ref{fig:allm-n}, but was not quite satisfactory in the low
$Q^2$ region.

The purpose of this note is to describe the results of a further
update of the ALLM parameterization, to be denoted ALLM97, where all
available published data of $F_2$, including the very low $x$, low
$Q^2$ data are used. As will be shown, this parameterization gives an
excellent description of all the data and can reproduce features like
the slope in $Q^2$ and in $x$, where all other parameterizations
fail. It can be used in the whole $x$ and $Q^2$ region above the
resonance region ($W^2>3$ GeV$^2$).

\section{The ALLM parameterization - a short recap}

  The proton structure function is assumed to have the form

\begin{equation}
   F_2(x,Q^2) = \frac{Q^2}{Q^2 + m_0^2}
                     \left(   F_2^{\mathcal P}(x,Q^2)
                            + F_2^{\mathcal R}(x,Q^2)
                     \right) ,
\end{equation}
where $m_0$ is the effective photon mass.  The functions
$F_2^{\mathcal P}$ and $F_2^{\mathcal R}$ are the contributions of the
Pomeron $\mathcal P$ or Reggeon $\mathcal R$ exchanges to the
structure function.  They take the form
\begin{equation}
\begin{array}{rcl}
   F_2^{\mathcal P}(x,Q^2)   &=&
       c_{\mathcal P}(t) x_{\mathcal P}^{a_{\mathcal P}(t)}
          (1 - x)^{b_{\mathcal P}(t)} ,  
\\
   F_2^{\mathcal R}(x,Q^2)   &=&
       c_{\mathcal R}(t) x_{\mathcal R}^{a_{\mathcal R}(t)}
          (1 - x)^{b_{\mathcal R}(t)} .
\end{array}
\label{eq:par}
\end{equation}  
The slowly varying function~$t$ is defined as
\begin{equation}
     t = \ln \left(  \frac{\ln \frac{Q^2 + Q^2_0}{\Lambda^2}}
                          {\ln \frac{Q^2_0}{\Lambda^2}}
             \right),
\end{equation}
where $\Lambda$ is the QCD scale and $Q^2_0$ is a parameter.

The two scaled variables $x_{\mathcal P}$ and $x_{\mathcal R}$ are
modified Bjorken--$x$ variables which include mass parameters
$m_{\mathcal P}$ and $m_{\mathcal R}$, interpreted as effective
Pomeron and reggeon masses:
\begin{equation}  
\begin{array}{rcl}
  \frac{1}{x_{\mathcal P}} &=& 1 + \frac{W^2 - M^2}{Q^2 + m_{\mathcal P}^2} ,
\\
  \frac{1}{x_{\mathcal R}} &=& 1 + \frac{W^2 - M^2}{Q^2 + m_{\mathcal R}^2} .
\end{array}
\end{equation}
where $M$ is the proton mass. The scale parameters
$m_0^2, m_{\mathcal P}^2, m_{\mathcal R}^2$, and $Q_0^2$, 
allow a smooth transition 
to $Q^2$ = 0 values. For large $Q^2$, $Q^2 \gg
m_{\mathcal P}^2, Q^2 \gg m_{\mathcal R}^2$, the scaled $x_{\mathcal
P}$ and $x_{\mathcal R}$ variables approach Bjorken $x$.

Four of the six parameters in equation~\ref{eq:par}, $c_{\mathcal R},
a_{\mathcal R}, b_{\mathcal R}$ and $b_{\mathcal P}$ increase with
$Q^2$ as
\begin{equation}
f(t) = f_1 + f_2 t^c
\end{equation}
while the remaining two, $c_{\mathcal P}$ and $a_{\mathcal P}$
decrease with $Q^2$ like:
\begin{equation}
g(t) = g_1 + (g_1 - g_2)\left[ \frac{1}{1 + t^d} - 1\right].
\end{equation}

There are altogether 23 parameters to be determined from a fit to the
data.  Note that about half of the parameters are needed for
the description of the low $W$ (high $x$) region where higher twist
effects are important.

The data used in the first fit (ALLM91) were all data available from the
pre--HERA era, which resulted in 694 data points. These included the
low energy data of $\sigma_{tot}(\gamma p)$~\cite{gp}, and the $F_2$
data of the SLAC~\cite{slac}, BCDMS~\cite{bcdms}, and NA28~\cite{na28}
collaborations. The best fit to the data had a $\chi^2$/ndf=0.98.

\begin{figure}[tbh]
\begin{center}
  \includegraphics [bb=40 162 517 636,width=\hsize,totalheight=7cm]
  {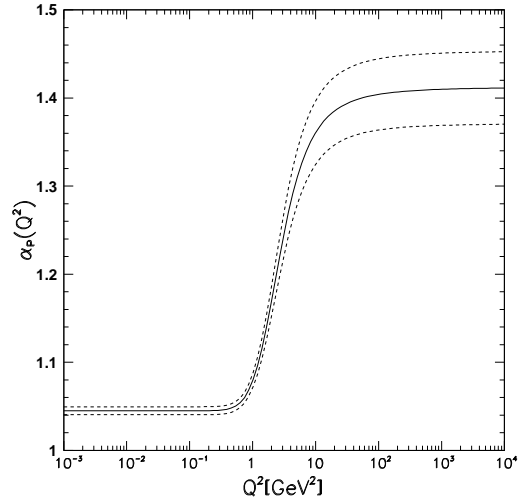}
\end{center}
\vspace{-.5cm}
\caption
 { The intercept of the Pomeron trajectory as function of $Q^2$, as
 obtained from the ALLM parameterization. The dotted line shows the
 uncertainty of the fit.
  }
\label{fig:ap-allm}
\end{figure}

As stated in the introduction and shown in figure~\ref{fig:allm-95},
the predictions of the parameterization ALLM91 to the first HERA data were in
agreement with the measurements at $Q^2$=0 and in the higher $Q^2$
region. Though some of the low $x$ parameters were constrained by the
NA28 measurements, the intermediate $Q^2$ region was not well
described, mainly due to the fact that the $Q^2$ dependence of the
Pomeron intercept, $\alpha_{\mathcal P}(Q^2)$, shown in
figure~\ref{fig:ap-allm}, came out too steep in the first fit.

The inclusion of the first HERA measurements as well as the
preliminary NMC data yielded the parameterization ALLM-N, which
produced a milder $Q^2$ transition of the Pomeron intercept, but 
fails to describe the latest low $x$ low $Q^2$ data. This prompted a
third attempt to determine the parameters including the latest
data, as described in the next section.

\section{The data sample for ALLM97}

The following data have been used for the present fit. All fixed
target photoproduction total cross section data were used together
with those of H1~\cite{h1gp} and ZEUS~\cite{zeusgp} in the HERA
region, a total of 228 data points. The $F_2$ structure function data
of the following fixed target collaborations: SLAC (211 points), BCDMS
(177), E665 (87)~\cite{e665}, and NMC (158)~\cite{nmc}. From the HERA
$ep$ collider we used the H1--94 data (193 points)~\cite{h194}, H1 low
$Q^2$ data (44)~\cite{h1lowq}, the ZEUS shifted vertex data
(36)~\cite{zeussv}, ZEUS--94 (188)~\cite{zeus94} and the very low
$Q^2$ ZEUS data measured with a beam pipe calorimeter (BPC) (34 data
points)~\cite{zeusbpc}. Altogether 1356 data points were used in the
fit resulting in a $\chi^2$/ndf=0.97. The contribution of each data
sample to the $\chi^2$ is given in table~\ref{tb:data}.
\begin{table}[h]
\centerline{\hbox{
\begin{tabular}{|l|c|c|}
\hline 
Data set & \# of points & $\chi^2$ \\ 
\hline\hline
$\gamma p$~\cite{gp},~\cite{h1gp},~\cite{zeusgp} & 228 & 262.3 \\ \hline
SLAC~\cite{slac} & 211 & 171.2 \\ \hline
BCDMS~\cite{bcdms} & 177 & 168.1 \\ \hline
E665~\cite{e665} & 87 & 95.7 \\ \hline
NMC~\cite{nmc}  & 158 & 142.0 \\ \hline
H1--94~\cite{h194} & 193 & 127.1 \\ \hline
H1(Low $Q^2$)~\cite{h1lowq} & 44 & 34.4 \\ \hline
ZEUS(SVX)~\cite{zeussv} & 36 & 26.7 \\ \hline
ZEUS--94~\cite{zeus94} & 188 & 253.8 \\ \hline 
ZEUS(BPC)~\cite{zeusbpc} & 34 & 17.3 \\ \hline \hline
Total & 1356 & 1298.7 \\ \hline
$\chi^2$/ND & & 0.97 \\ \hline
\end{tabular}
}}
\bf\caption
{\it
Data used in the ALLM97 fit, with the $\chi^2$ contribution of each set.
}
\label{tb:data}
\end{table}

\section{Results}

\begin{figure}[h]
\begin{center}
  \includegraphics [bb=40 162 517 636,width=\hsize,totalheight=7cm]
  {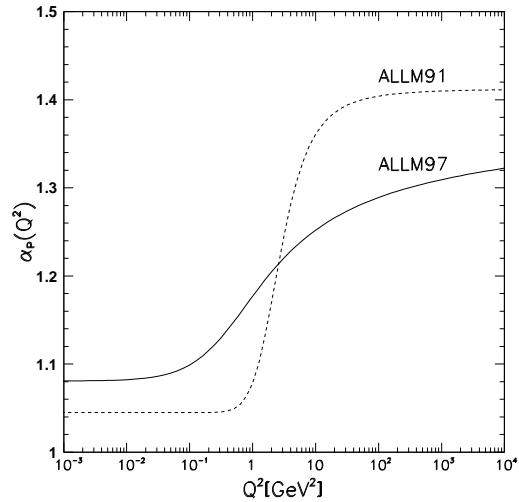}
\end{center}
\vspace{-.5cm}
\caption
 { The intercept of the Pomeron trajectory as function of $Q^2$, as
 obtained from the ALLM97 (full line) and ALLM91 (dotted line)
 parameterizations.  }
\label{fig:ap-97-91}
\end{figure}

The resulting parameter values of the ALLM97 fit are compared in
table~\ref{tb:par} to their values from the ALLM91 fit.
\begin{table}[h]
\centerline{\hbox{
\begin{tabular}{|l||c|c|} \hline 
Parameter & ALLM91 & ALLM97 \\ \hline \hline
$m_0^2$(GeV$^2$) & 0.30508 & 0.31985 \\ \hline
$m_{\mathcal P}^2$(GeV$^2$) & 10.676  & 49.457  \\ \hline
$m_{\mathcal R}^2$(GeV$^2$) & 0.20623 & 0.15052 \\ \hline
$Q_0^2$(GeV$^2$) & 0.27799 & 0.52544 \\ \hline
$\Lambda^2$(GeV$^2$) & 0.06527 & 0.06527  \\ \hline
$c_{{\mathcal P}1}$ & 0.26550 & 0.28067 \\ \hline
$c_{{\mathcal P}2}$ & 0.04856 & 0.22291 \\ \hline
$c_{{\mathcal P}3}$ & 1.04682 & 2.1979 \\ \hline
$a_{{\mathcal P}1}$ & -0.04503 & -0.0808 \\ \hline
$a_{{\mathcal P}2}$ & -0.36407 & -0.44812 \\ \hline
$a_{{\mathcal P}3}$ & 8.17091 & 1.1709 \\ \hline
$b_{{\mathcal P}1}$ & 0.49222 & 0.36292 \\ \hline
$b_{{\mathcal P}2}$ & 0.52116 & 1.8917 \\ \hline
$b_{{\mathcal P}3}$ & 3.5515 & 1.8439 \\ \hline
$c_{{\mathcal R}1}$ & 0.67639 & 0.80107 \\ \hline
$c_{{\mathcal R}2}$ & 0.49027 & 0.97307 \\ \hline
$c_{{\mathcal R}3}$ & 2.66275 & 3.4942 \\ \hline
$a_{{\mathcal R}1}$ & 0.60408 & 0.58400 \\ \hline
$a_{{\mathcal R}2}$ & 0.17353 & 0.37888 \\ \hline
$a_{{\mathcal R}3}$ & 1.61812 & 2.6063 \\ \hline
$b_{{\mathcal R}1}$ & 1.26066 & 0.01147 \\ \hline
$b_{{\mathcal R}2}$ & 1.83624 & 3.7582 \\ \hline
$b_{{\mathcal R}3}$ & 0.81141 & 0.49338 \\ \hline
\end{tabular}
}}
\bf\caption
{\it
The parameter values in the old (ALLM91) and new (ALLM97) parameterization.
}
\label{tb:par}
\end{table}
The biggest difference can be seen in the value of the scale parameter
of the Pomeron which increased by almost a factor of 5. This increase
affects the shape of the transition region in the low $x$ low $Q^2$
region. This can be seen in figure~\ref{fig:ap-97-91} where the
dependence of the Pomeron intercept on $Q^2$ is plotted for the old
(ALLM91) and the new (ALLM97) parameterization.
The latter allows for an early start of the transition from the soft
to the hard regime. Note that in the present fit the intercept at
$Q^2$=0 was fixed to the Donnachie--Landshoff (DL)~\cite{dl} value
since the total photoproduction measurements in the HERA region do not
allow a precise determination of this value.

\subsection{$F_2$ as function of $Q^2$}

\begin{figure}[tbh]
\begin{center}
  \includegraphics [bb=6 102 536 702,width=\hsize,totalheight=20cm]
  {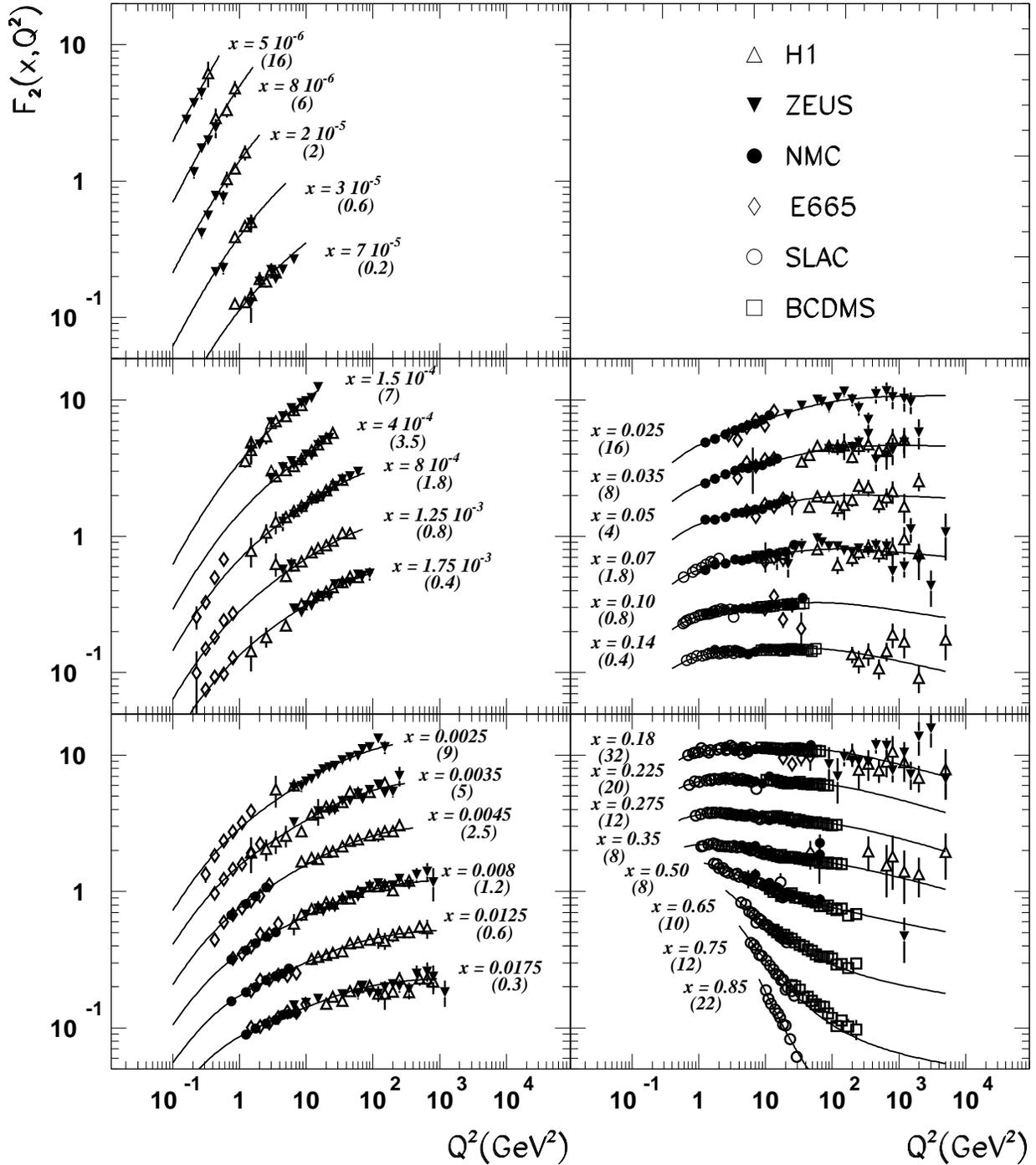}
\end{center}
\vspace{-.5cm}
\caption
{ The dependence of the proton structure function, $F_2(x,Q^2)$, on
$Q^2$ for fixed $x$ values as indicated in the figure. For display
purposes, the structure function values have been scaled at each $x$
by the factor shown in brackets under the $x$ values. The curves are
the results of the ALLM97 parameterization.  }
\label{fig:f2all}
\end{figure}

The $F_2$ data~\cite{hawar} used in the fit are displayed in
figure~\ref{fig:f2all} as function of $Q^2$ for fixed $x$ intervals,
together with the results of the ALLM97 parameterization. One sees the
well known scaling violation behaviour of the data, being positive for
low $x$ values and turning negative in the high
$x$ region. 

The curves in the figure are the results of the ALLM97
parameterization and are seen to go through most of the data in the whole
($x,Q^2$) region. This reflects the good $\chi^2$ obtained from the fit. 

\subsection{$\sigma_{tot}$ as function of $Q^2$}

\begin{figure}[h]
\begin{center}
  \includegraphics [bb=63 215 510 651,width=\hsize,totalheight=20cm]
  {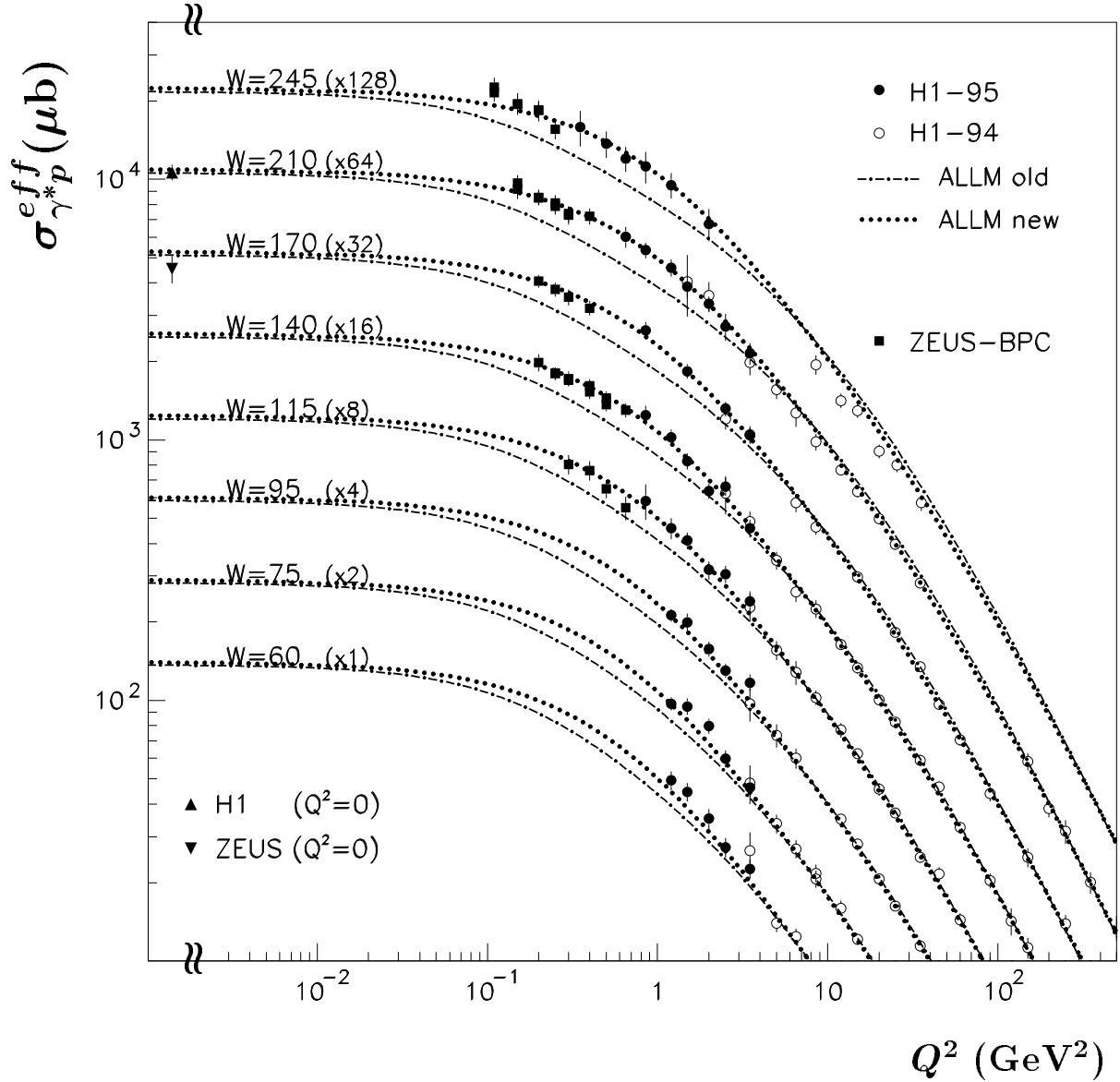}
\end{center}
\vspace{-.5cm}
\caption
 { The effective virtual photon-proton cross section as function of
 $Q^2$ for fixed $W$ intervals. The full lines are the results of the
 ALLM97 parameterization and the dotted lines are those of the ALLM-N
 ones.  }
\label{fig:h1-eff}
\end{figure}

The H1 collaboration~\cite{h1lowq} presented their results together
with those of the ZEUS BPC data as an effective virtual photon-proton
cross section, $\sigma^{eff}_{\gamma^*p}$, as function of $Q^2$ for
fixed $W$ intervals. The effective cross section is given by
$\sigma^{eff}_{\gamma^*p} = \sigma_T + \epsilon \sigma_L$, where
$\sigma_T$ and $\sigma_L$ are the cross sections for transverse and
longitudinally polarized virtual photons and $\epsilon$ is the ratio
of longitudinal to transverse flux.  The data are displayed in
figure~\ref{fig:h1-eff}. Since in the HERA kinematic region $\epsilon
\approx $ 1, the data are compared to the old (ALLM-N) and new
(ALLM97) parameterization of $\sigma_{tot}(\gamma^* p)$. The ALLM97
parameterization gives a good description of the data. Also shown are
the two points at $Q^2$=0 which are also well described by the new
parameterization.

\subsection{$\sigma_{tot}$ as function of $W^2$}

\begin{figure}[h]
\begin{center}
  \includegraphics [bb=18 96 567 753,width=\hsize,totalheight=20cm]
  {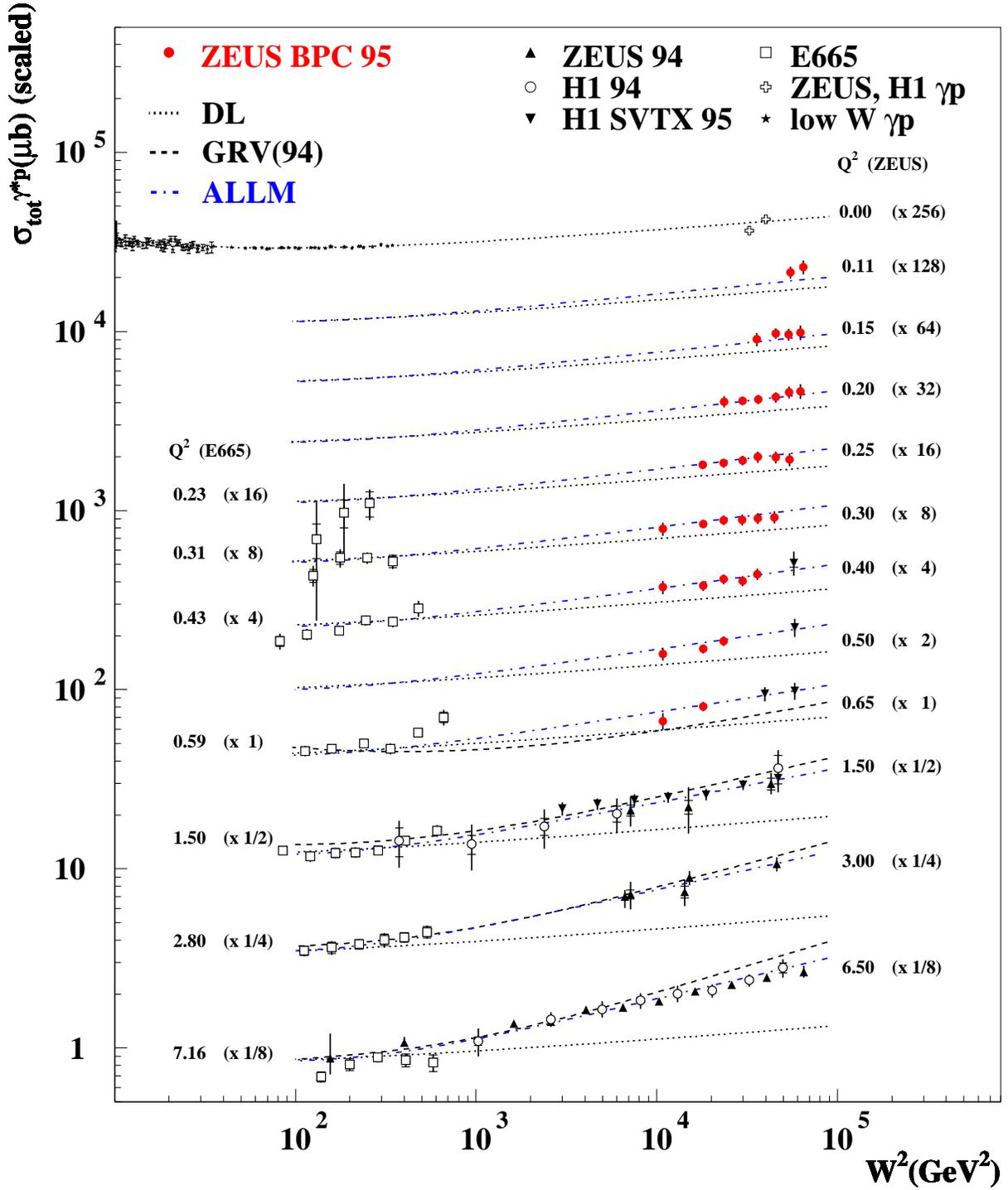}
\end{center}
\vspace{-.5cm}
\caption
{ The total virtual photon-proton cross section as function of $W^2$
for low $Q^2$ data, scaled with the factors indicated in brackets for
display purposes. The curves are the expectations of different
parameterizations.  }
\label{fig:zeus-lowq}
\end{figure}

The $F_2$ data can be converted to $\sigma_{tot}(\gamma^* p)$ using
the relation
\begin{equation}
\sigma_{tot}(\gamma^* p) = \frac{4\pi^2\alpha}{Q^2(1-x)}
\frac{Q^2 + 4M^2x^2}{Q^2}F_2(W^2,Q^2).
\end{equation}
The $F_2$ data at low $Q^2$, starting as low as $Q^2$=0.11 GeV$^2$,
are shown in figure~\ref{fig:zeus-lowq} in the form of
$\sigma_{tot}(\gamma^* p)$ together with the real photon total cross
sections. The data are compared to expectations of different
parameterizations.
While the Donnachie--Landshoff (DL) parameterization agrees well with
$\sigma_{tot}(\gamma p)$, its predictions are lower than the data once
$Q^2 \ne$ 0, with the disagreement increasing with $Q^2$. The
GRV~\cite{grv} parameterization is plotted starting at $Q^2$ = 0.65
GeV$^2$, where it lies below the data, while at higher $Q^2$ values
the predictions are above the data. The ALLM97 parameterization agrees
with the data at all $Q^2$ values.

\begin{figure}[h]
\begin{center}
  \includegraphics [bb=16 76 534 739,width=\hsize,totalheight=20cm]
  {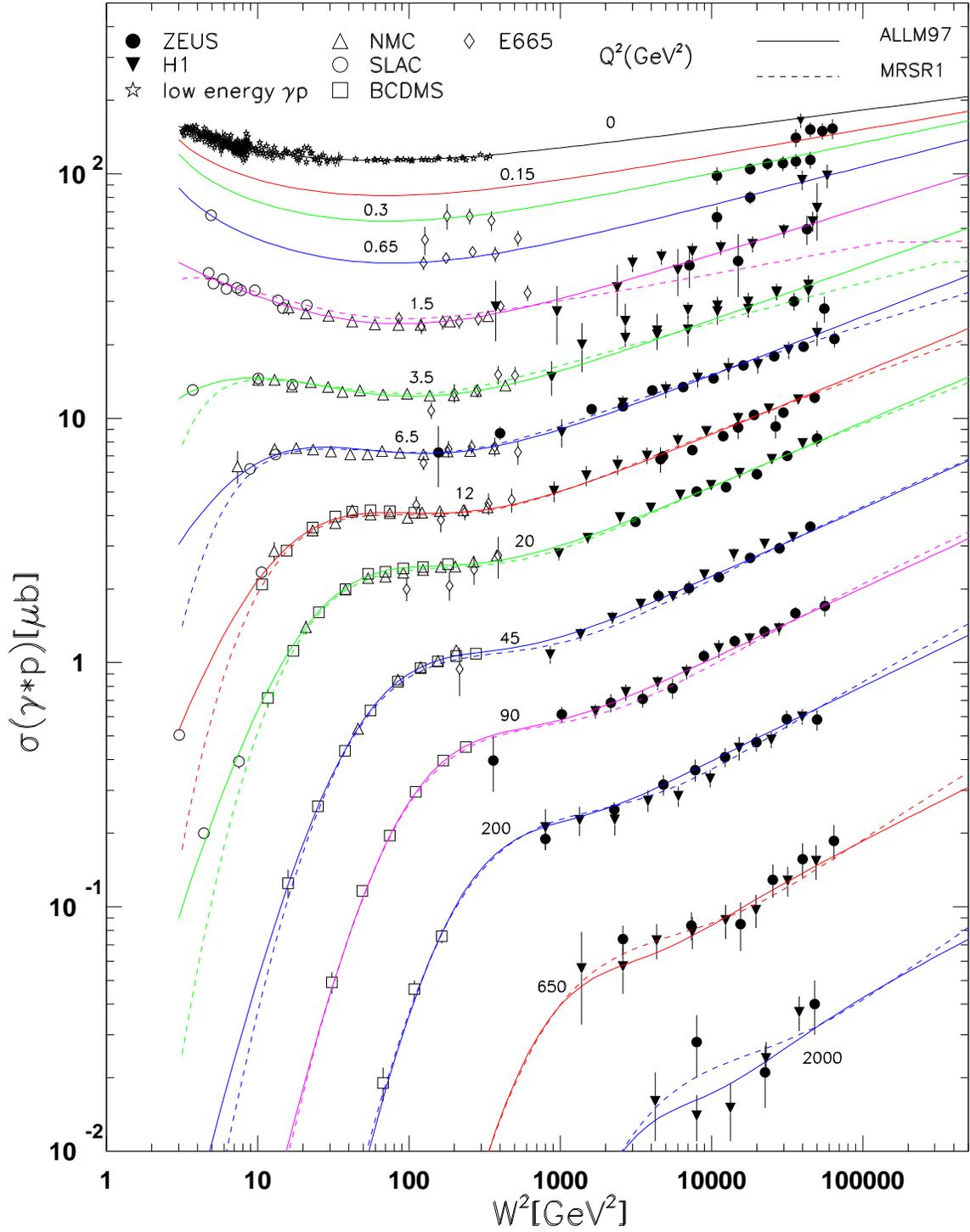}
\end{center}
\vspace{-.5cm}
\caption
{The total $\gamma^* p$ cross section as function of $W^2$, for
different $Q^2$ values. The curves are the expectations of the ALLM97
parameterization (full line) and those of the MRSR1 one (dotted line).  }
\label{fig:allm-mrsr1}
\end{figure}

The good agreement of the ALLM97 parameterization in the whole
accessible kinematic region at present can be seen in
figure~\ref{fig:allm-mrsr1} where the total cross section is plotted in
the range $0 \le Q^2 \le 2000$ GeV$^2$.
It gives a good description of the data at high as well as at low $W^2$. 

In order to compare the results of the ALLM97 parameterization to that
of a recent QCD evolution type of parameterization, we show in the
same figure also the MRSR1~\cite{mrsr1} parameterization which is
valid for $Q^2 >$ 1.25 GeV$^2$.  The two parameterizations agree well
with each other for $Q^2 \geq$ 10 GeV$^2$, while at lower $Q^2$ values
the MRSR1 parameterization has a shallower $W^2$ dependence and thus
is lower than the data.

\section{The transition region}

The different $W^2$ behaviour of the high $Q^2$ data and at $Q^2$=0
prompted the measurements of deep inelastic $ep$ reactions in the low
$Q^2$ region in order to find where the transition takes place. We
will look at two ways of studying this question. One is by looking at
the change in the slope of $F_2$ with respect to $\ln Q^2$ and the
other is to study $d\ln F_2/d\ln x$, both of which are discussed
below.

\subsection{$dF_2/d\ln Q^2$ as function of $x$ for some $Q^2$ values}

The scaling violation of $F_2$ is expected to increase as $x$ gets
smaller according to QCD. This feature is also borne out by the data
as shown in figure~\ref{fig:f2all}. One can quantify this by looking
at the change of the slope of $F_2$ with respect to $\ln Q^2$ for
different $x$ values. Once the non--perturbative processes take over,
as expected at low $Q^2$, one should see a change in the slope. The
place where the change occurs would indicate the transition from soft
to hard physics.

\begin{figure}[h]
\begin{center}
  \includegraphics [bb=43 166 514 704,width=\hsize,totalheight=8cm]
  {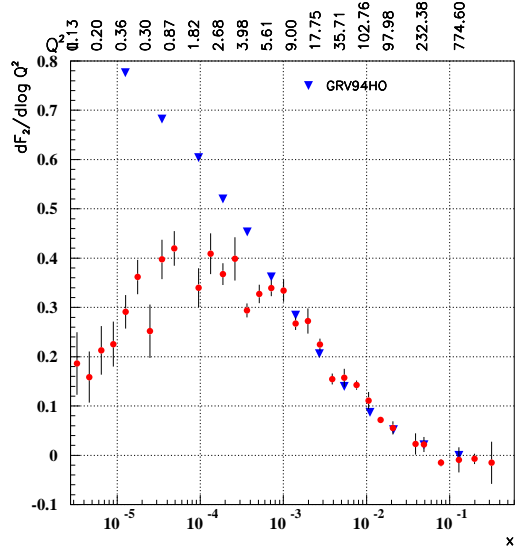}
\end{center}
\vspace{-.5cm}
\caption
{ The slope of $F_2$ with respect to $\ln x$ as function of $x$ for
$Q^2$ values as indicated in the figure. The round dots are the HERA
data and the triangle symbols are the results of the GRV94HO
parameterization. }
\label{fig:hera-grv}
\end{figure}

The distribution of $dF_2/d\ln Q^2$ as function of $x$~\cite{allen} is
shown for the HERA data in figure~\ref{fig:hera-grv} for $Q^2$ values
ranging from about 1000 GeV$^2$ down to 0.13 GeV$^2$. Some values are
given at the top of the figure. The HERA data include the H194, H1 low
$Q^2$, ZEUS94, ZEUS shifted vertex and ZEUS BPC data.  As expected,
the slope rises as $x$ decreases down to $x \approx 10^{-4}$. However
for lower $x$ there is a change in the tendency of the slope which
becomes smaller as $x$ decreases. This happens at $Q^2$ values of
about 1-2 GeV$^2$. Note that as $x$ decreases also $Q^2$ decreases.
In the same figure we plot for comparison the expectation of the GRV
parameterization. This parameterization starts its evolution at $Q^2
\approx$ 0.4 GeV$^2$.  While the parameterization shows the same
features as the data for $Q^2 >$ 5 GeV$^2$, it continues to rise with
$x$ also below $x = 10^{-4}$, contrary to the data. Judging from the
GRV distribution, the turnover point starts as high as $Q^2 \approx$ 4
GeV$^2$.

\begin{figure}[h]
\begin{center}
  \includegraphics [bb=39 179 512 705,width=\hsize,totalheight=8cm]
  {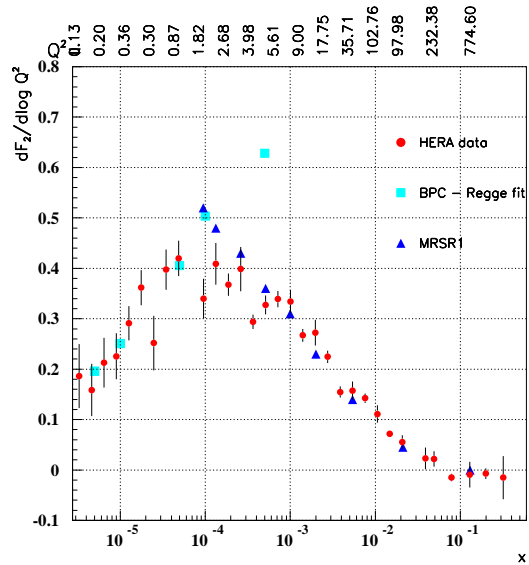}
\end{center}
\vspace{-.5cm}
\caption
{ The slope of $F_2$ with respect to $\ln x$ as function of $x$ for
$Q^2$ values as indicated in the figure. The round dots are the HERA
data, the triangle symbols are the results of the MRSR1
parameterization and the full squares come from a Regge fit to the BPC
data. }
\label{fig:hera-mrs-regge}
\end{figure}

Another way of trying to find the turning point is to compare the same
data to another QCD evolution type parameterization and to a Regge
based one. This is done in figure~\ref{fig:hera-mrs-regge}, where the
data are compared to the results of the MRSR1 parameterization and to
the expectations from a Regge fit which was done~\cite{bpc-regge} to
the BPC data. The MRSR1 parameterization start its evolution at $Q^2$
= 1.25 GeV$^2$, where it is higher than the data until about $Q^2
\approx $ 3-4 GeV$^2$ from whereon it follows the data. The Regge fit 
starts from the lowest $Q^2$ point and agrees with the data up to about
$Q^2$ = 1 GeV$^2$, but continues to rise at higher $Q^2$ values
contrary to the data. The QCD and Regge results cross at $Q^2
\approx$ 2 GeV$^2$. One could thus conclude from here that the transition 
region is in the region of 1 - 3 GeV$^2$.

\begin{figure}[h]
\begin{center}
  \includegraphics [bb=9 104 543 759,width=\hsize,totalheight=14cm]
  {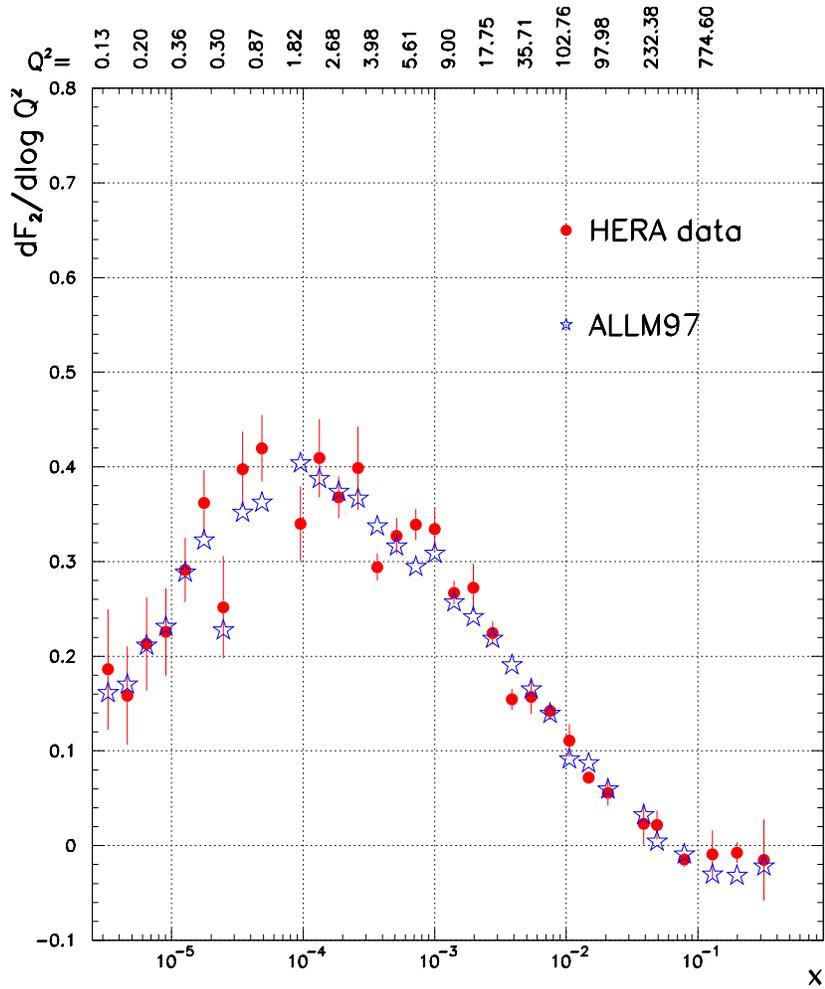}
\end{center}
\vspace{-.5cm}
\caption
{ The slope of $F_2$ with respect to $\ln x$ as function of $x$ for
$Q^2$ values as indicated in the figure. The full dots are the HERA
data and the open symbols are the results of ALLM97. }
\label{fig:hera-slopes}
\end{figure}

Finally, in figure~\ref{fig:hera-slopes} the data are compared to the
ALLM97 parameterization.  One observes good agreement between the
parameterization and the data. Thus one does not need to use two
different parameterizations to describe the low $Q^2$ soft and the
high $Q^2$ hard regimes. ALLM97 gives a good parameterization of both
regimes.

\subsection{ $d\ln F_2/d\ln x$ as function of $Q^2$}

The slope of $d\ln F_2/d\ln x$ can be related to the Pomeron
intercept. In the low $x$ region, $F_2$ is behaving like
$x^{-\lambda}$, where $\lambda$ is a function of $Q^2$. Since for
fixed $Q^2$, $W^2 \sim x^{-1}$, $\sigma_{tot}(\gamma^* p) \sim
W^{2\lambda}$ and therefore $\lambda = \alpha_{\pom}$ -
1. Therefore, measuring $d\ln F_2/d\ln x$ as function of $Q^2$ is
equivalent to measuring $\lambda$. When making this interpretation it
is crucial to choose the right $x$ region from which the slope is
determined.

\begin{figure}[h]
\begin{center}
  \includegraphics [bb=11 129 539 663,width=\hsize,totalheight=20cm]
  {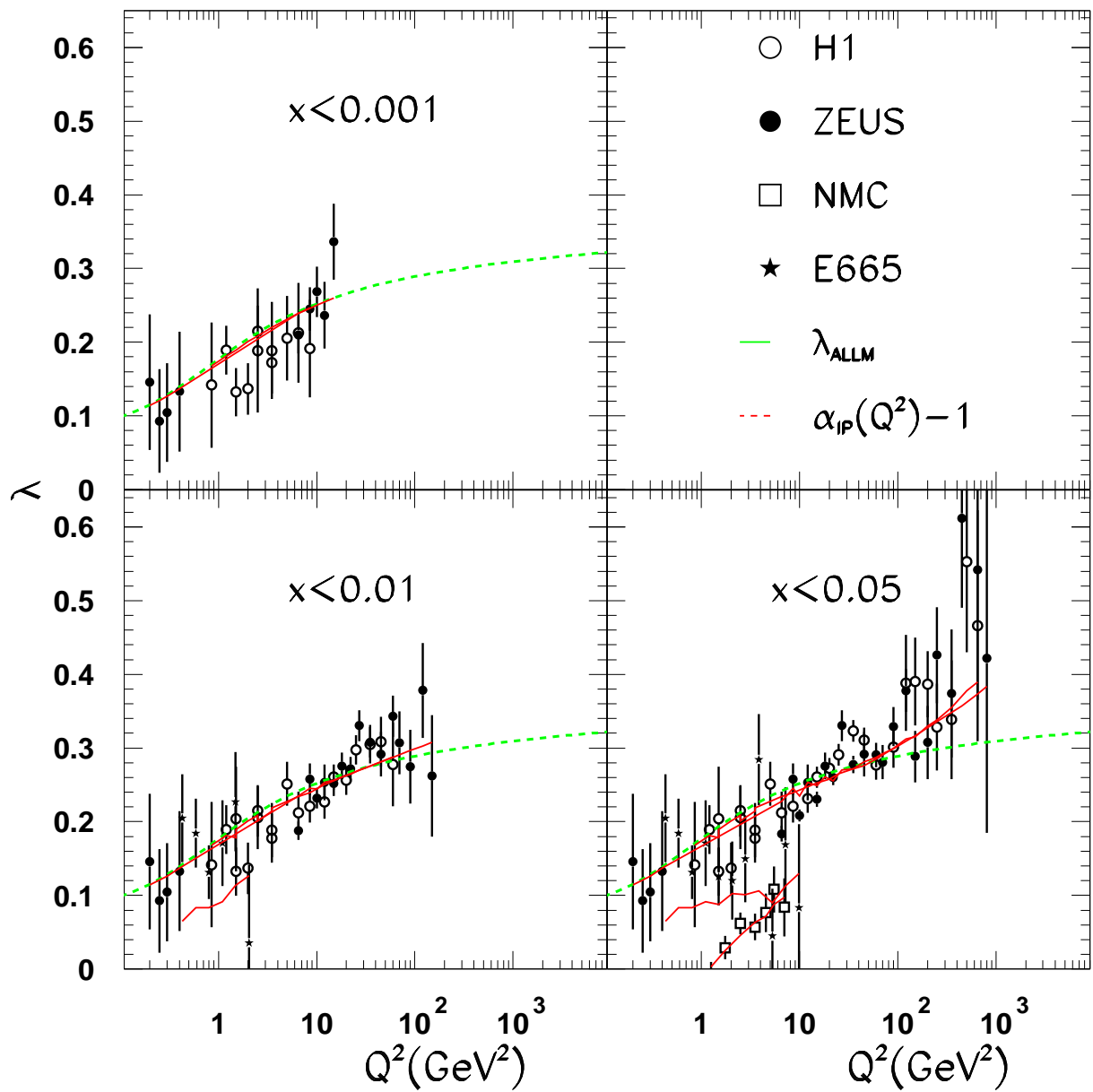}
\end{center}
\vspace{-.5cm}
\caption
{ The slope $\lambda$ as function of $Q^2$ for different $x$
cuts, as indicated in the figure. The dashed line is $\alpha_P$ - 1 as
calculated from the ALLM97 parameterization, while the full line is the value
of $\lambda$ as obtained from the parameterization in the same way as from
the data.
 }
\label{fig:lplot-data}
\end{figure}

This last point is demonstrated in figure~\ref{fig:lplot-data} where
the slope $\lambda$ as function of $Q^2$ is shown for different $x$
cuts, as indicated in the figure, for the separate data sets of each
experiment. Note that in order to obtain $\lambda$ at a fixed $Q^2$, a
minimum of 4 data points at different $x$ values were required. The
full line is $\alpha_{\pom}$ - 1 as calculated from the ALLM97
parameterization, where $\alpha_{\pom}$ is the Pomeron intercept, and
the dashed line gives the value of $\lambda$ obtained from the
parameterization in the same procedure as that applied to the data.
As one sees, as long as one uses data in the very low $x$ region, $x <
10^{-3}$, one can correlate the measured slope $\lambda$ with the
Pomeron intercept. For larger $x$ cuts $\lambda$ may not be always a
good estimate of $\alpha_{\pom}$-1, as shown for example for $x <$
0.05. The NMC data in the low $Q^2$ region do not reach very low $x$
values, their lowest being $x$=0.0045, and thus produce low
$\lambda$ results. The HERA data in the region $Q^2 >$ 200 GeV$^2$
overestimate the Pomeron intercept by determining the slope in a
narrow $x$ region, typically $0.01 < x < 0.05$. These effects are
reproduced by the ALLM97 parameterization.

\section{Summary and conclusions}

The ALLM parameterization has been updated by using all the published
data to determine its parameters leading to ALLM97. A very good
description of the data in the whole ($x,Q^2$) kinematic region is
obtained, including the $Q^2$ = 0 photoproduction measurements down to
$W^2$ = 3 GeV$^2$ and the low $Q^2$ and low $x$ region where the
transition from soft to hard processes is observed in the data.

We have hereby demonstrated that it is possible to find a functional
form which describes the data in the whole of the kinematical
region. Such a parameterization has many practical applications and in
addition allows to study features of the data which are helpful for
understanding the interplay between soft and hard processes.

\section*{Acknowledgments}
 
We would like to thank our colleagues Eugene Levin and Uri Maor for
their continuous support in this study.  We also thank Vladimir
Chekelyan and Bernd Surrow for their help with some of the figures.
The critical reading of the text by Guenter Wolf is highly
appreciated.

\noindent This work was partially supported by the German--Israel 
Foundation (GIF) and the Israel Science Foundation.




\end{document}